\documentclass[prx,floatfix,twocolumn,tightenlines,nofootinbib,superscriptaddress]{revtex4-1}
\usepackage{amsmath,amsthm,amssymb}
\usepackage{graphicx}
\usepackage{times}
\usepackage{hyperref}
\hypersetup{
colorlinks=true,
linkcolor=red,
citecolor=red,}
\usepackage{textcomp}
\usepackage{xcolor}
\usepackage{enumerate}
\usepackage{multirow}
\usepackage{tabularx}
\usepackage{dcolumn}
\usepackage[mathscr]{euscript}
\usepackage{mathtools}
\usepackage{textcomp}
\usepackage{calrsfs}
\usepackage[separate-uncertainty=true]{siunitx}
\usepackage{soul}
\usepackage{comment}

\newcommand{\bra}[1] {\langle #1 |}
\newcommand{\ket}[1] {| #1 \rangle}
\newcommand{\braket}[2] {\langle #1 | #2 \rangle}

\newcommand{\vs}[1]{{\boldsymbol{\rm #1}}}

\begin{document}

\title[Assisted Macroscopic Quantumness]{Assisted Macroscopic Quantumness}

\author{Farid Shahandeh}
\email{Electronic address: shahandeh.f@gmail.com}
\affiliation{Department of Physics, Swansea University, Singleton Park, Swansea SA2 8PP, United Kingdom}
\affiliation{Centre for Quantum Computation and Communication Technology, School of Mathematics and Physics, University of Queensland, St Lucia, Queensland 4072, Australia}
\author{Martin Ringbauer}
\affiliation{Institut f\"{u}r Experimentalphysik, Universit\"{a}t Innsbruck, Technikerstra\ss{}e 25, 6020 Innsbruck, Austria}
\affiliation{Scottish Universities Physics Alliance (SUPA), Institute of Photonics and Quantum Sciences, School of Engineering and Physical Sciences, Heriot-Watt University, Edinburgh EH14 4AS, UK}
\author{Massimiliano Proietti}
\affiliation{Scottish Universities Physics Alliance (SUPA), Institute of Photonics and Quantum Sciences, School of Engineering and Physical Sciences, Heriot-Watt University, Edinburgh EH14 4AS, UK}
\author{Fabio Costa}
\affiliation{Centre for Engineered Quantum Systems, School of Mathematics and Physics, University of Queensland, St Lucia, Queensland 4072, Australia}
\author{Austin P. Lund}
\affiliation{Centre for Quantum Computation and Communication Technology, School of Mathematics and Physics, University of Queensland, St Lucia, Queensland 4072, Australia}
\author{Francesco Graffitti}
\affiliation{Scottish Universities Physics Alliance (SUPA), Institute of Photonics and Quantum Sciences, School of Engineering and Physical Sciences, Heriot-Watt University, Edinburgh EH14 4AS, UK}
\author{Peter Barrow}
\affiliation{Scottish Universities Physics Alliance (SUPA), Institute of Photonics and Quantum Sciences, School of Engineering and Physical Sciences, Heriot-Watt University, Edinburgh EH14 4AS, UK}
\author{Alexander Pickston}
\affiliation{Scottish Universities Physics Alliance (SUPA), Institute of Photonics and Quantum Sciences, School of Engineering and Physical Sciences, Heriot-Watt University, Edinburgh EH14 4AS, UK}
\author{Dmytro Kundys}
\affiliation{Scottish Universities Physics Alliance (SUPA), Institute of Photonics and Quantum Sciences, School of Engineering and Physical Sciences, Heriot-Watt University, Edinburgh EH14 4AS, UK}
\author{Timothy C. Ralph}
\affiliation{Centre for Quantum Computation and Communication Technology, School of Mathematics and Physics, University of Queensland, St Lucia, Queensland 4072, Australia}
\author{Alessandro Fedrizzi}
\affiliation{Scottish Universities Physics Alliance (SUPA), Institute of Photonics and Quantum Sciences, School of Engineering and Physical Sciences, Heriot-Watt University, Edinburgh EH14 4AS, UK}

\begin{abstract}
It is commonly expected that quantum theory is universal, in that it describes the world at all scales. Yet, quantum effects at the macroscopic scale continue to elude our experimental observation. This fact is commonly attributed to decoherence processes affecting systems of sufficiently large number of constituent subsystems leading to an effective macro-scale beyond which a quantum description of the whole system becomes superfluous. Here, we show both theoretically and experimentally that the existence of such a scale is unjustifiable from an information-theoretic perspective. We introduce a variant of the Wigner's friend experiment in which a multiparticle quantum system is observed by the friend. The friend undergoes rapid decoherence through her interactions with the environment. In the usual version of this thought experiment, decoherence removes the need for Wigner to treat her as a quantum system. However, for our variant we prove theoretically and observe experimentally that there exist partitions of the subsystems in which the friend is entangled with one of the particles in assistance with the other particle, as observed by Wigner. Importantly, we show that the friend is indispensable for the entanglement to be observed. Hence Wigner is compelled to treat the friend as part of a larger quantum system. By analyzing our scenario in the context of a quantum key distribution protocol, we show that a semi-classical description of the experiment is suboptimal for security analysis, highlighting the significance of the quantum description of the friend.
\end{abstract}

\maketitle

\section{Introduction}
Quantum theory provides one of the most successful descriptions of the physical world, yet its mathematical structure allows for counter-intuitive phenomena, such as superposition of states, nonlocality, and entanglement. In microscopic experiments these features are now observed on a regular basis and if quantum mechanics is assumed to be a \emph{universal} theory, it should, in principle, be possible to observe them with arbitrarily large objects. Although experiments are making great progress in this direction~\cite{Friedman2000,Julsgaard2001,Raimond2001,Leggett2002,Marshall2003,DeMartini2012,Sciarrino2013,Arndt2014,Zarkeshian2017,Lee2017}, it remains an open question---known as the \emph{macroscopicity problem}---whether certain quantum effects can indeed be observed at the level of our everyday experience, as epitomized by Schr\"{o}dinger's famous cat~\cite{Schrodinger1935}.

A commonly conjectured roadblock for macroscopic quantum effects is quantum decoherence related to the size of the considered system~\cite{JoosBook,Zurek2003}. 
Loosely speaking, according to some (rather intuitive) definitions of macroscopic quantumness wherein the size of the system (and thus the number of its constituent elements) is a decisive parameter~\cite{Raeisi2011,Lee2011,Frowis2012,Sekatski2014-1,Sekatski2014-2,Laghaout2014,Farrow2015,Frowis2018}, the larger the system is the harder it is to isolate it from interactions with the environment. Such interactions, in turn, destroy the coherence of the system, hence, no quantum property of a macroscopic, i.e. \emph{sufficiently large}, system can be observed unless via extremely high-precision measurements~\cite{Raeisi2011,Sekatski2014-1}. In other words, it is assumed that there exists a ``scale'' beyond which physical systems become ``macroscopic'' and can be analyzed without any reference to the quantum formalism.

\begin{figure}[h]
\centering
  \includegraphics[width=0.8\columnwidth]{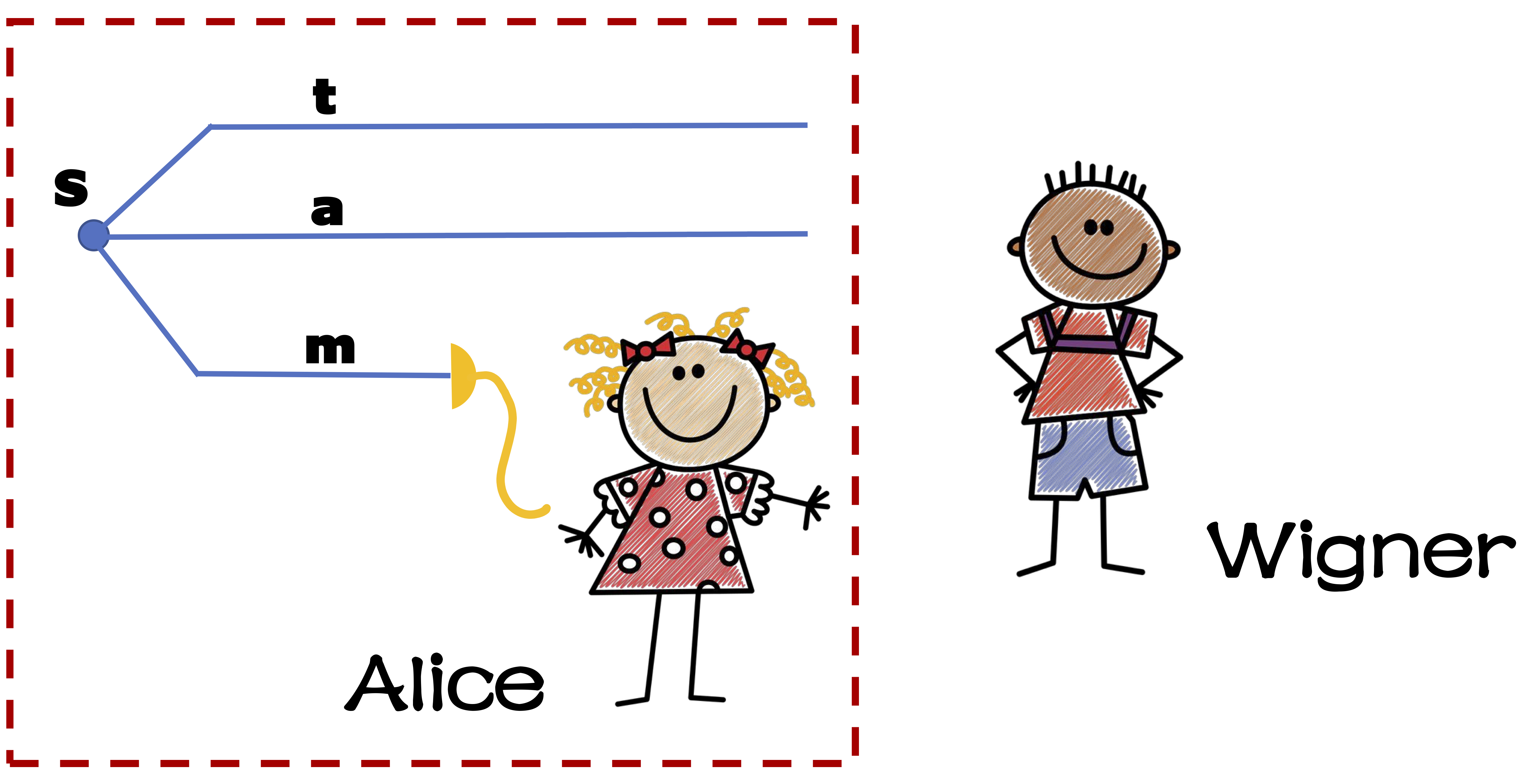}
  \caption{\textbf{The schematic of our protocol.} A source $\vs{S}$ produces an entangled pair in modes ${\vs t}$ and ${\vs m}$ and an ancilla photon in mode ${\vs a}$. Photons ${\vs a}$ and ${\vs t}$ then interact to produce the state of Eq.~\eqref{initial}. The state undergoes partial decoherence before the qubit in mode $\vs m$ is measured by the macroscopic observer Alice, who also undergoes decoherence. Wigner, outside Alice's laboratory, concludes that Alice and any of the two qubits $\vs a$ or $\vs t$ are jointly entangled with the other qubit, while there is no entanglement between the qubits $\vs a$ and $\vs t$ alone.}
  \label{fig:Scheme}
\end{figure}

In this contribution we challenge this view by showing in theory and experiment that, with the assistance of a second system, a macroscopic system (in the sense of being made up by a very large number of subsystems) can be proven to be entangled even after complete decoherence. We set the scene as in a Wigner's friend experiment, with the crucial difference that the observer is not assumed to preserve quantum coherence, and analyze Wigner's perspective, see Fig.~\ref{fig:Scheme}. In our variant, Wigner's friend is in possession of two particles, labeled $\vs a$ and $\vs t$. We find that, even in the presence of decoherence and regardless of its dynamics, the joint subsystem of Wigner's friend and particle $\vs t$ exhibits entanglement with particle $\vs a$, while there is no entanglement in any pair of subsystems alone. Consequently, as far as the information content of Wigner's friend is concerned, she constitutes a quantum system in assistance with particle $\vs t$ regardless of her size. More specifically, Wigner has to use the quantum formalism to describe the entanglement in this particular partition, independent of our choice of interpretation of quantum theory. Hence, there seems to be no escape from the conclusion that Wigner's friend, despite being a macroscopic observer by any sensible definition, is an \emph{informationally indispensable} part of the quantum system. We provide a link between such a quantum property of macroscopic observers and the security of quantum key distribution (QKD) protocols in terms of the distinguishability of the friend's macroscopic states and argue that a semi-classical analysis of this situation is sub-optimal, thus reflecting the necessity of both the \emph{presence} and a quantum treatment of Wigner's friend.

\section{Results}

\subsection{The standard Wigner's friend scenario}\label{sec:OrigWigner}
In Wigner's seminal thought experiment~\cite{Wigner1995} a spin-$1/2$ system, labeled $\vs a$, is initially in the superposition state $\ket{+}_{\vs a}{=}(\ket{0}_{\vs a} {+} \ket{1}_{\vs a})/\sqrt{2}$, where $\ket{0}$ stands for ``spin up'' and $\ket{1}$ for ``spin down''. Wigner's friend (whom we call Alice from now on) then measures the spin and randomly obtains either the result ``up'' or ``down'' in each run of the experiment. Wigner, who is outside the closed laboratory, must---under the assumption that quantum mechanics is universal in the sense that it can formally describe all physical systems, whether they display non-classical features or not---describes the whole laboratory as a pure quantum state evolving according to the Schr\"{o}dinger equation into the state $\ket{\chi}_{\vs{aA}}^{(\vs{W})}=(\ket{0}_{\vs a}\ket{\text{``up''}}_{\vs A} {+} \ket{1}_{\vs a}\ket{\text{``down''}}_{\vs A})/\sqrt{2}$ where the subindex $\vs A$ denotes Alice.
Here and in the following, we use the superscripts within parentheses to specify the observer who assigns the quantum state to the systems in their possession, to be specific, $\vs{A}$ and $\vs{W}$ indicate the observers Alice and Wigner, respectively. The details of how this state has been obtained, such as coupling to Alice's sense organs~\cite{Deutsch1985}, are irrelevant for the argument. Strikingly, since the formalism of quantum mechanics makes no distinction between small and large systems, we now find the macroscopic Alice as part of the \emph{entangled} quantum state $\ket{\chi}_{\vs{aA}}^{(\vs{W})}$.

A further counter-intuitive feature of this situation is revealed as Wigner aims to verify his state assignment through an interference experiment on Alice and her quantum system~\cite{Wigner1995}. Quantum theory predicts that, if Wigner upon measuring the qubit $\vs{a}$ in a superposition-basis, finds it in state $\ket{+}_{\vs a}$, he will observe Alice being in a superposition state $(\ket{\text{``up''}}_{\vs A} {+} \ket{\text{``down''}}_{\vs A})/\sqrt{2}$, due to their initial entanglement. This not only contradicts our classical experience (the macroscopicity problem), but also highlights the ambiguity in distinguishing the observed system from the observer, also known as the system-apparatus cut, in quantum mechanics~\cite{Bacciagaluppi2013}. The latter is central to the infamous and yet unresolved \emph{measurement problem}~\cite{Brukner2015MeasurementProblem}, which has received a lot of renewed attention recently~\cite{Brukner2018,Frauchiger2016SingleWorld,Proietti2019}.

It is commonly suggested that the practical resolution of both the macroscopicity and the measurement problems is that decoherence makes it extremely difficult, if not impossible, to realize such an experiment. Indeed, a dynamical analysis of the decoherence process reveals that the size of the system plays a significant role, inducing the idea that there might be a scale beyond which decoherence renders a quantum analysis of the system irrelevant~\cite{JoosBook,Zurek2003}. In the standard Wigner's friend thought experiment, this would imply that, due to decoherence, the state of the total system from Wigner's perspective reduces to $(\ket{0}_{\vs a}\bra{0} {\otimes} \hat{\tau}_{\vs A} + \ket{1}_{\vs a}\bra{1} {\otimes} \hat{\upsilon}_{\vs A})/2$, with $\hat{\tau}_{\vs A}$ and $\hat{\upsilon}_{\vs A}$ representing Alice's states after decoherence. The precise form of Alice's states is irrelevant to recognizing that this joint state after decoherence will not show interference in Wigner's thought experiment. Thus, in this case, the placement of the cut becomes irrelevant and the problematic incompatibility between Alice's and Wigner's observations supposedly vanishes in the face of decoherence. Although there are critics of this position~\cite{Pearle1998,JoosBook,Adler2003,Rosaler2016}, it is argued that decoherence at least retrieves classicality at a macroscopic level in an operational sense~\cite{Bacciagaluppi2013}.

\subsection{The modified Wigner's friend scenario}
We now construct a modified Wigner's friend experiment that is not subject to this argument, illustrated in Fig.~\ref{fig:Scheme}. In this experiment, macroscopic decoherence \emph{does not} diminish the informational content of the macroscopic system and does not imply classicality of the whole system, because it is unable to destroy the entanglement between all parties. Consequently, the size of the macro-system (or its environment) has no effect on the stored information within the system rendering the details of the decoherence dynamics irrelevant. As a result, decoherence effects cannot be used to justify the proclamation of a scale beyond which the quantum formalism becomes redundant.

\subsubsection{The friend's perspective}
Consider a source of photons that produces a three-mode entangled state of the form
\begin{equation}
\label{initial}
\ket{\Phi}_{\vs{atm}} = \left(\ket{\Psi_+}_{\vs{at}}\ket{0}_{\vs m} {+} \ket{\Psi_-}_{\vs{at}}\ket{1}_{\vs m}\right)/\sqrt{2},
\end{equation}
where $\ket{\Psi_\pm}_{\vs {at}} {=} (\ket{01}_{\vs {at}}\pm\ket{10}_{\vs {at}})/\sqrt{2}$.
Here, the subscripts $\vs{a}$, $\vs{t}$, and $\vs{m}$ label the three modes.
In contrast to the standard Wigner's friend experiment, we now allow the state to interact with the environment inside the lab, which, from Alice's point of view, leads to dephasing of the state within the partition $(\vs{at}|\vs{m})$. In other words, we assume decoherence relative to the system that will eventually be measured by the macroscopic observer, while making the empirically justified assumption that the coherence between the remaining qubit systems can be maintained. This can happen, for instance, via the subsystem $\vs{m}$ interacting unitarily with {\it inaccessible} ancillae modes (environment), as is considered in the standard approaches to open systems. We emphasize, however, that the details of the decoherence process is not relevant to the argument to follow. The resulting state after decoherence is given by
\begin{equation}
\label{atm}
\hat{\varrho}_{\vs {atm}}^{(\vs A)} {=} \left(\ket{\Psi_+}_{\vs{at}}\bra{\Psi_+} {\otimes} \ket{0}_{\vs m}\bra{0} {+} \ket{\Psi_-}_{\vs{at}}\bra{\Psi_-} {\otimes} \ket{1}_{\vs m}\bra{1}\right)/2.
\end{equation}
 A crucial observation at this point is that the state assigned by Alice features no entanglement between any pair of qubits, i.e.\ within partitions $(\vs a|\vs t)$, $(\vs a|\vs m)$, and $(\vs m|\vs t)$. Nonetheless, Alice finds that her state is entangled within partitions $(\vs{am}|\vs t)$ and $(\vs a|\vs{tm})$, implying that two of the qubits are entangled with the other qubit in these particular partitions. To see this, consider the negativity as a convex (distillable) entanglement monotone, defined as~\cite{Vidal2002}
\begin{equation}
\label{EN}
E_N[\hat{\varrho};({\vs X|\vs Y})]:=\frac{\|\hat{\varrho}^{\mathsf{T}_{\vs X}}\|_1 -1}{2},
\end{equation}
where $^{\mathsf{T}_{\vs X}}$ denotes partial transposition with respect to subsystem $\vs X$ and $\|\cdot\|_1$ is the $\ell_1$-norm, i.e. the sum of the absolute values of the eigenvalues. A direct calculation gives $E_N[\hat{\varrho}_{\vs{atm}}^{(\vs A)};({\vs{am}|\vs t})]{=}E_N[\hat{\varrho}_{\vs{atm}}^{(\vs A)};({\vs a|\vs{tm}})]{=}1/2$, which is not surprising to Alice, since, despite the state being mixed, the system under her investigation is ultimately quantum. Importantly, Alice's conclusion regarding the entanglement and thus the quantumness of the state $\hat{\varrho}_{\vs{atm}}^{(\vs A)}$ is independent of her interpretation of quantum theory. Alice then performs a measurement on the qubit in mode $\vs m$ to update her knowledge of the ancilla and target ($\vs a$ and $\vs t$) subsystems.

\begin{figure}[h]
\centering
  \includegraphics[width=1\columnwidth]{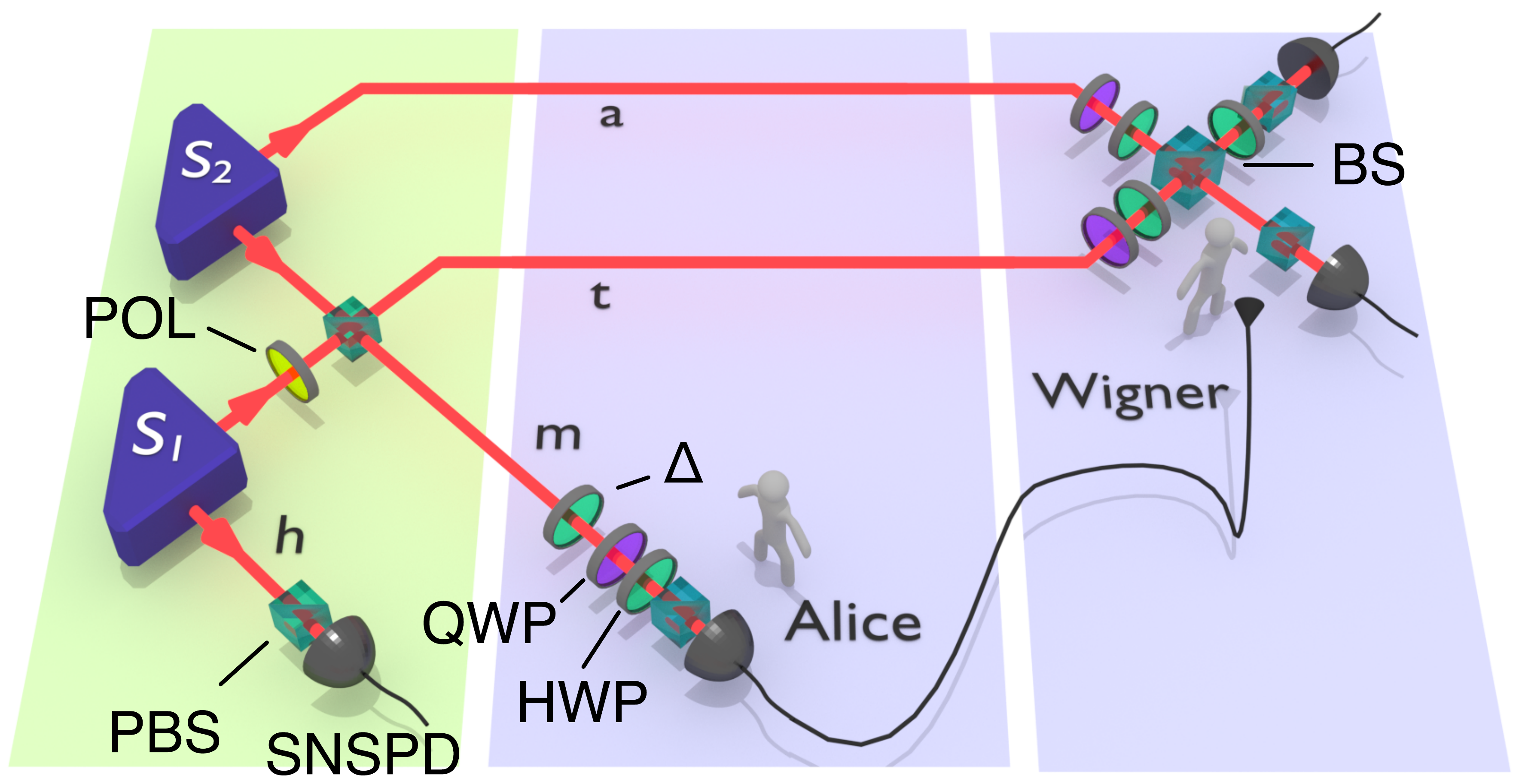}
  \caption{\textbf{Experimental setup}. The three-mode entangled state of Eq.~\eqref{initial} is generated in modes $\vs m, \vs a$, and $\vs t$ using two Sagnac-based spontaneous parametric downconversion sources~\cite{Fedrizzi2007Sagnac}, of which $S_1$ is set to prepare the entangled state $\ket{\Psi_-}$ and $S_2$ the separable state $\ket{+}\ket{0}$ using polarizers (POL). The photons are then combined in a type-I fusion gate~\cite{Browne2005} implemented through a polarizing beam splitter (PBS). One of the 4 emitted photons is detected in the auxiliary mode $\vs h$ to provide a heralding signal for the state preparation. In order to simulate environmental decoherence, the so-generated state can be continuously dephased up to the state of Eq.~\eqref{atm}, by imparting phase flips on mode $\vs m$ to an appropriate fraction of the experimental runs using a half-waveplate (HWP), denoted $\Delta$. Alice then measures the qubit in mode $\vs m$ using a set of HWP and quarter-waveplates (QWP) and PBS before detecting the photon using superconducting nanowire detectors (SNSPD). In order to measure the witnesses of Eq.~\eqref{witness}, arbitrary Bell-basis measurements on modes $\vs{at}$ are achieved using non-classical interference in a 50/50 beamsplitter (BS) combined with HWP and QWP rotations and coincidence detection. See Materials and Methods for details.}
  \label{fig:Setup}
\end{figure}

\subsubsection{Wigner's perspective}
From Wigner's perspective, outside the closed laboratory, the situation is  different. Following the quantum prescription, Wigner assigns the state $\ket{\overline{\Phi}}_{\vs{atmA}} =( \ket{\Psi_+}_{\vs{at}}\ket{0}_{\vs m}\ket{\xi}_{\vs A} + \ket{\Psi_-}_{\vs{at}}\ket{1}_{\vs m}\ket{\zeta}_{\vs A})/\sqrt{2}$. Here, $\ket{\xi}_{\vs A}$ and $\ket{\zeta}_{\vs A}$ are the memory states of Alice encoding her measurement result, where by assuming $|\braket{\xi}{\zeta}_{\vs A}| {=} \epsilon$ with $\epsilon\geq 0$ we allow Alice to err in her inference of the qubit in mode $\vs m$ to be in the $0$ or $1$ state. Without requiring any detailed description of the decoherence dynamics of the systems within the laboratory and merely by taking into account the decoherence of the initial state, and that Alice undergoes decoherence, Wigner assigns to the remaining ancilla-target-Alice system after Alice's measurement a state of the form
\begin{equation}\label{atAW}
\hat{\varrho}_{\vs{atA}}^{(\vs W)} = \left(\ket{\Psi_+}_{\vs{at}}\bra{\Psi_+}{\otimes} \hat{\tau}_{\vs A} + \ket{\Psi_-}_{\vs{at}}\bra{\Psi_-}{\otimes} \hat{\upsilon}_{\vs A}\right)/2.
\end{equation}

We assume that the system in mode $\vs m$ is destroyed by Alice's measurement, as is typically the case in photonic experiments such as the one in Fig.~\ref{fig:Setup}, but non-destructive measurements (e.g.\ in spin systems) would lead to an equivalent description. Again, the states $\hat{\tau}_{\vs A}$ and $\hat{\upsilon}_{\vs A}$ represent Alice's state after decoherence, where for our purpose it is enough that $\hat{\tau}_{\vs A}$ and $\hat{\upsilon}_{\vs A}$ encode some information about Alice's perception of her measurement outcomes. In contrast to the decoherence version of the standard Wigner's friend thought experiment, we now show that in the present experiment it {\it does} matter where the system-apparatus cut is placed. By calculating the negativity of the state in Eq.~\eqref{atAW} we obtain
\begin{equation}\label{WEN}
E_N\![\hat{\varrho}_{\vs{atA}}^{(\vs W)};\!({\vs{aA}|\vs t})] =  E_N\![\hat{\varrho}_{\vs{atA}}^{(\vs W)};\!({\vs a|\vs{tA}})] = \frac{\| \hat{\tau}_{\vs A} {-} \hat{\upsilon}_{\vs A}\|_1}{4}.
\end{equation}
Wigner must therefore conclude, independent of his of interpretation of quantum theory, that:
(i)~Alice is part of a large entangled state within partitions $(\vs{aA}|\vs t)$ and $(\vs a|\vs{tA})$;
(ii)~the amount of entanglement within both partitions is determined by the distinguishability of Alice's memory states;
(iii) the entanglement \emph{is not} merely due to qubits, as Alice is \emph{necessary} for obtaining any entanglement. In other words, if Alice is disregarded by tracing her out, there is no entanglement within the remaining system $(\vs a|\vs t)$. Importantly, both of the states $\hat{\varrho}_{\vs{atm}}^{(\vs A)}$ and $\hat{\varrho}_{\vs{atA}}^{(\vs{W})}$ are entangled on equal footings, since there is nothing within the theory that makes an informational difference between the $\vs m$-qubit and Alice making them operationally equivalent. Consequently, the placement of the system-apparatus cut is critical in our experiment in the sense that the observations made by Wigner and Alice remain incompatible, even under full decoherence. Only when Alice is admitted a quantum treatment can the entanglement of the joint system be revealed.

\subsection{Differences to the conventional analysis}
We now briefly discuss the crucial differences in the analysis and interpretation of our scenario compared to the conventional Wigner's friend experiment. There has recently been an increasing interest in the modified Wigner's friend scenarios, both in theory~\cite{Brukner2018,Frauchiger2016SingleWorld} and experiment~\cite{Proietti2019,Bong2019}. These works are primarily devoted to the disconnect between Wigner’s perception of reality and that of the friend. In contrast, our aim is to investigate the extent to which the assumption of the existence of a “macro-scale” beyond which a quantum description becomes redundant is valid.

Here, we have extended the macroscopic system ``Alice" of the conventional proposals to an even larger macroscopic system $\vs{Aa}$ (that is, Alice plus the qubit $\vs{a}$). This effectively debilitates the decoherence mechanism's ability to reach ``classicality'', since it only leads to a loss of coherence between Alice and the microscopic subsystems. Technically, while decoherence destroys the entanglement in the partition $(\vs{A}|\vs{at})$, this is insufficient to achieve macroscopic ``classicality'', since the macroscopic system $\vs{aA}$ (which is larger than Alice) remains entangled to the qubit $\vs{t}$. At the same time, there is no entanglement in the partition $(\vs{a|t})$, implying that the observed entanglement is \emph{not} merely due to the microscopic subsystems $\vs{a}$ and $\vs{t}$.

At this point one may be tempted to think of Alice as simply holding a piece of (classical) information, which identifies one or the other entangled quantum state of subsystem $\vs{at}$.
We emphasize that this requires one to refer to the individual subsystem $\vs{A}$ which, in turn, means working with either the bipartition $(\vs{A}|\vs{at})$ or the tripartition $(\vs{A}|\vs{a}|\vs{t})$.
While this view justifies some kind of classical-quantum description of the whole system across the partitions ($\vs{A|at}$) or ($\vs{A|a|t}$), it fails to provide such a description for the partition ($\vs{aA|t}$), which includes the larger macroscopic system $\vs{aA}$ of interest as an entity here, because the classical theories required to describe $\vs{A}$ do not possess the structure to accommodate entanglement. As such, the entanglement across the partition ($\vs{aA|t}$) can only be captured when endowing the macroscopic system $\vs{aA}$ with a fully quantum description.
Indeed, conditioning on Alice's information is a way to transfer the entanglement from the partition ($\vs{aA|t}$) to the partition ($\vs{a|t}$). This emphasizes the crucial role Alice's information plays for the entanglement of the whole system. Consequently, we come to the conclusion that there are micro-macro systems that show quantum features, even in the presence of decoherence.

\subsection{Witnessing the entanglement involving the macro-object}
Assuming the universality of quantum theory, Alice and Wigner are now able to verify the entanglement of their states in many ways, including using entanglement witnessing techniques. Wigner can verify the entanglement of the state in Eq.~\eqref{atAW} by performing measurements on the ancilla and target qubits and asking Alice about her measurement results. The latter restriction to asking Alice only classical questions ensures that the experiment is well within the realm of current technology. Specifically, the two witnesses
\begin{equation}\label{witness}
\begin{split}
&\hat{W}_{1}{=}\ket{\Phi_+}_{\vs {at}}\bra{\Phi_+}^{\mathsf{T}_{\vs t}}{\otimes} \ket{\text{``up''}}_{\vs A}\bra{\text{``up''}},\\
&\hat{W}_{2}=\ket{\Phi_+}_{\vs {at}}\bra{\Phi_+}^{\mathsf{T}_{\vs t}}{\otimes} \ket{\text{``down''}}_{\vs A}\bra{\text{``down''}},
\end{split}
\end{equation}
with $\ket{\Phi_\pm}_{\vs {at}} {=} (\ket{00}_{\vs {at}}\pm\ket{11}_{\vs {at}})/\sqrt{2}$, can be used for this purpose whenever there is nonzero distinguishability between the states $\hat{\tau}_{\vs A}$ and $\hat{\upsilon}_{\vs A}$ in the basis $\{\ket{\text{``up''}}_{\vs A},\ket{\text{``down''}}_{\vs A}\}$, as we show in the Materials and Methods. A given state $\hat{\varrho}^\star$ is entangled within partitions $(\vs{aA}|\vs t)$ and $(\vs a|\vs{tA})$ if ${\rm Tr}\hat{\varrho}^\star\hat{W}_1 {<}0$, or ${\rm Tr}\hat{\varrho}^\star\hat{W}_2 {<} 0$. This witnessing procedure is operationally equivalent to Wigner asking Alice about her memory and measuring $\ket{\Phi_+}_{\vs{at}}\bra{\Phi_+}^{\mathsf{T}_{\vs t}}$ for the ancilla-target two-qubit subsystem. Conditioning on Alice's answer being ``up'' or ``down'', Wigner evaluates the expectation values ${\rm Tr}\hat{\varrho}^\star\hat{W}_1$ and ${\rm Tr}\hat{\varrho}^\star\hat{W}_2$, where a negative value for either of them implies the entanglement within the partitions $(\vs a|\vs{tA})$ and $(\vs{aA}|\vs t)$. In a quantitative sense, the value by which the witnessing inequality is violated also puts a lower bound on the amount of entanglement present in the quantum state~\cite{Brandao2005,Shahandeh2017}. Alice may use the same recipe to witness the entanglement of her state by substituting ``asking Alice'' with ``measurements in the computational basis on mode $\vs m$''.

\begin{figure*}[ht!]
  \includegraphics[width=\textwidth]{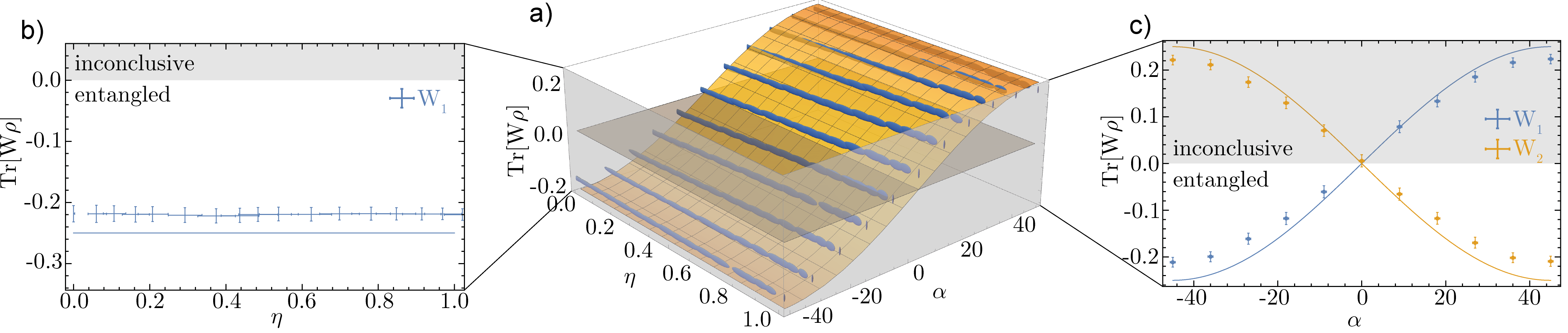}
  \caption{\textbf{Experimental results.} \textbf{a)} Measured expectation values $\langle \hat{W}_1\rangle$ for the witness $\hat{W}_1$ as a function of the amount of dephasing $\eta$ and the distinguishability of Alice's measurement results as measured by $\alpha$. The orange plane represents the theory prediction, the gray plane indicates the boundary below which $\hat{W}_1$ certifies the presence of entanglement in the partitions $(\vs a|\vs{tA})$ and $(\vs t|\vs{aA})$. The size of the blue data points in the 3-dimensional plot correspond to 3-sigma statistical uncertainty obtained from a Monte Carlo routine simulating the Poissonian counting statistics. \textbf{b)} Measured expectation values $\langle \hat{W}_1\rangle$ for different amounts of dephasing in the limiting case of perfect distinguishability ($\alpha=45^\circ$). \textbf{c)} Measured expectation values $\langle \hat{W}_1\rangle$ (blue) and $\langle \hat{W}_2\rangle$ (orange) for different distinguishability of Alice's measurements in the case of full dephasing ($\eta = 1$). The gray shaded area indicates the region where the witness is inconclusive, while the white area is where either witness certifies entanglement. All error bars represent 3-sigma statistical uncertainty regions.}
  \label{fig:Results3D}
\end{figure*}

We implemented the modified Wigner's friend scenario in a photonic experiment as depicted in Fig.~\ref{fig:Setup} using two independent photon-pair sources and a type-I fusion gate~\cite{Browne2005}. We measured the witnesses in Eq.~\eqref{witness}, and verified Wigner's conclusion about the entanglement of his state $\hat{\varrho}_{\vs{atA}}^{(\vs W)}$. In order to consider the effects of decoherence, we experimentally implement a single-qubit dephasing channel $\rho\mapsto(1-\frac{\eta}{2})\rho + \frac{\eta}{2} \sigma_z \rho \sigma_z$ on mode ${\vs m}$ before Alice's measurement via random phase flips, see Fig.~\ref{fig:Setup}. This allows us to continuously tune the state before Alice's measurement from the pure state of Eq.~\eqref{initial} to the fully decohered state of Eq.~\eqref{atm}, see Methods for the experimental details. The results of Fig.~\ref{fig:Results3D} show that the observed entanglement is independent of the strength $\eta$ of the decoherence applied to the state. This clearly shows that Alice remains an indispensable part of an entangled state even under full decoherence and independent of any micro-macro distinction.
Additionally, we considered the effects of imperfect distinguishability of Alice's results, by having her implement a measurement described by the non-orthogonal projectors $\{\cos(\pi /4 \pm \alpha)\ket{0} + \sin(\pi /4\pm \alpha)\ket{1}\}$ on the qubit in mode ${\vs{m}}$, see Fig.~\ref{fig:Setup}. For $\alpha{=}\pm\pi/4$ this implements the ideal measurement of qubit ${\vs m}$ in the $\{\ket{0},\ket{1}\}$ basis, while for $\alpha{=}0$ the measurement reveals no information about the qubit's polarization state. This measurement thus enables us to study the case where Alice's memory does not allow for a perfect identification of the state in mode $\vs m$, where the distinguishability of her measurement results is given by $\| \hat{\tau}_{\vs A} {-} \hat{\upsilon}_{\vs A}\|_1 = |\sin[2\alpha]|$. To measure the entanglement witness of Eq.~\eqref{witness}, from Wigner's perspective, we measured the two qubits in modes $\vs a$ and $\vs t$ and asked Alice for her observed results. The results in Fig.~\ref{fig:Results3D} indicate that either of the witness $\hat{W}_1$ or $\hat{W}_2$ certifies the presence of entanglement in the partitions $(\vs a|\vs{tA})$ and $(\vs t|\vs{aA})$ for all non-zero values of distinguishability of Alice's measurement results (i.e.\ $\alpha\neq 0$).

\subsection{Practical consequences}
\begin{figure}[h]
\centering
  \includegraphics[width=0.9\columnwidth]{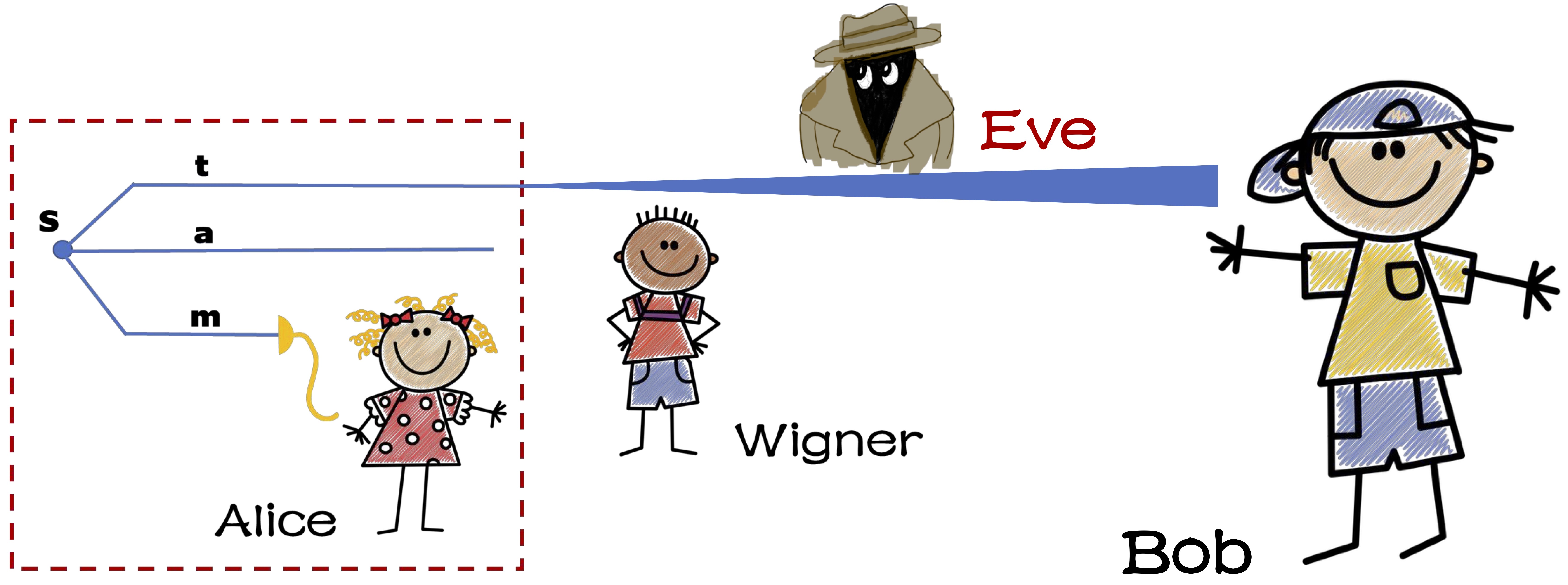}
  \caption{\textbf{A QKD scenario.} Wigner may send the ancilla qubit $\vs{t}$ over a distance to Bob to distribute a secure quantum key. The security of the key against Eve is thereby directly determined by the distinguishability of Alice's macroscopic states.}
  \label{fig:QKDScheme}
\end{figure}

These observations have intriguing practical consequences, which highlight the importance of a quantum description of Alice through the states $\hat{\tau}_{\vs A}$ and $\hat{\upsilon}_{\vs A}$. Suppose that Wigner wants to securely distribute quantum keys with a third party, Bob, in a distant laboratory. To accomplish this he would send one of the qubits, say qubit ${\vs a}$ to Bob and keep the subsystem of target-Alice, see Fig.~\ref{fig:QKDScheme}. Now assume that an eavesdropper, Eve, has access to the information that flowed to the environment of Alice; hence, Eve potentially holds the purification of the state $\hat{\varrho}_{\vs {atA}}^{(\vs W)}$ shared between Wigner and Bob. This state has exactly the form of a \emph{private} state, with $\ket{\Psi_\pm}_{\vs {at}}$ being the \emph{key} and $\hat{\tau}_{\vs A}$ and $\hat{\upsilon}_{\vs A}$ being the \emph{shield} parts~\cite{Horodecki2005}. Its density matrix in the computational basis of the qubits, i.e., $\{\ket{00}_{\vs{at}},\ket{01}_{\vs{at}},\ket{10}_{\vs{at}},\ket{11}_{\vs{at}}\}$, can be written as
\begin{equation}
\hat{\varrho}_{\vs {atA}}^{(\vs W)}=
\begin{pmatrix}
\hat{\xi}_{\vs A} & 0 & 0 & \hat{\zeta}_{\vs A} \\ 
0 & 0 & 0 & 0 \\ 
0 & 0 & 0 & 0 \\ 
\hat{\zeta}_{\vs A} & 0 & 0 & \hat{\xi}_{\vs A}
\end{pmatrix},
\end{equation} 
where $\hat{\xi}_{\vs A}{=}(\hat{\tau}_{\vs A} {+} \hat{\upsilon}_{\vs A})/4$ and $\hat{\zeta}_{\vs A}{=}(\hat{\tau}_{\vs A} {-} \hat{\upsilon}_{\vs A})/4$. This is the matrix representation of a Bell state with nonzero elements being replaced with operators corresponding to the subsystem Alice. According to Refs.~\cite{Horodecki2005,Horodecki2006}, Wigner and Bob will obtain one arbitrarily secure key bit even though an eavesdropper holds the purification of the state if the trace norm of the off-diagonal blocks tends to $1/2$, i.e., $\|\hat{\zeta}_{\vs A}\|_1{\rightarrow} 1/2$. This is related to the fact that for a Bell state the corresponding elements equal $1/2$.

In our experiment the trace-norm condition is directly related to the distinguishability of Alice's states; the more distinguishable her states are, the more distillable the entanglement of $\hat{\varrho}_{\vs {atA}}^{(\vs W)}$, and the more secure the key will be. As we have shown previously in Eq.~\eqref{WEN}, it follows that $0 {\leqslant} E_N[\hat{\varrho}_{\vs {atA}}^{(\vs W)};({\vs a|\vs{tA}})]{=}\|\hat{\zeta}_{\vs A}\|_1{\leqslant} 1/2$. The minimum (maximum) distillable entanglement is obtained when $\hat{\tau}_{\vs A}$ and $\hat{\upsilon}_{\vs A}$ are completely indistinguishable (perfectly distinguishable). In particular, perfectly distinguishable states imply perfect security of the distilled key. In this scenario, Eve can tell which of the states $|\Psi_+\rangle_{\vs {at}}$ or $|\Psi_-\rangle_{\vs {at}}$ Wigner and Bob share, because she has access to the purification of the shared state. However, she can never learn their key bit. As Horodecki \textit{et al}~\cite{Horodecki2005} state, ``In a sense, Eve can hold one bit of information but it is the wrong bit of information. Such a situation is impossible classically (or with pure quantum states held by Alice [Wigner] and Bob).''

The advantage of our complete quantum description over a semi-classical approach, where Alice's states are replaced by probabilities, becomes clear when measuring the amount of entanglement obtainable by Wigner and Bob. Suppose the correspondence between classical and quantum propositions is given by $\text{``up?''} {\leftrightarrow} \hat{\Pi}_{\rm u}$ and $\text{``down?''} {\leftrightarrow} \hat{\Pi}_{\rm d}$, where $\{\hat{\Pi}_{\rm u},\hat{\Pi}_{\rm d}\}$ is a POVM so that $\hat{\Pi}_{\rm u}{+}\hat{\Pi}_{\rm d}{=}\hat{I}$. Hence, the analogous classical probability distributions associated with the quantum states are $p_{\rm u/d}{=}{\rm Tr}\hat{\Pi}_{\rm u/d}\hat{\tau}_{\vs A}$ and $q_{\rm u/d}{=}{\rm Tr}\hat{\Pi}_{\rm u/d}\hat{\upsilon}_{\vs A}$. The semi-classical approach would give exactly the same statistical results {\it only if} the distance between classical distributions of Alice equals their respective quantum distinguishability, i.e., $\frac{1}{2}\sum_{x={\rm u,d}}|p_x-q_x| {=} \|p-q\|_1 {=} 2\|\hat{\zeta}_{\vs A}\|_1$.
It is, however, well known that this is not generally true, as~\cite{Nielsen} 
\begin{equation}
\|\hat{\zeta}_{\vs A}\|_1 = \frac{1}{2}\max_{\{\hat{\Pi}_m\}} \|\tilde{p}-\tilde{q}\|_1,
\end{equation}
where the maximization is over \emph{all possible} POVMs, $\tilde{p}_{m} {=} {\rm Tr}\hat{\Pi}_{m}\hat{\tau}_{\vs A}$, and $\tilde{q}_{m} {=} {\rm Tr}\hat{\Pi}_{m}\hat{\upsilon}_{\vs A}$.
Consequently, $\|p-q\|_1 {\leqslant} 2\|\hat{\zeta}_{\vs A}\|_1 {=} 2E_N[\hat{\varrho}_{\vs {atA}}^{(\vs W)};({\vs a|\vs{tA}})]$, implying that the semi-classical approach only results in a lower bound on the amount of distillable entanglement. As discussed in Materials and Methods, this is in line with the observation that the violation of a witnessing inequality only puts a lower bound on the amount of entanglement within the quantum state~\cite{Brandao2005,Shahandeh2017}. This shows that an optimal security analysis of such protocols requires a quantum treatment with more general measurements of macroscopic systems, as a semi-classical analysis may be insufficient.

Finally, we show explicitly within the Materials and Methods that the situation described in our example is generic, i.e., from Wigner's perspective, Alice's knowledge of the constituent elements of an ensemble, even though limited, implies that she is part of an entangled quantum system independent of the entanglement measure used.


\section{Discussion}
We have introduced a variant of the seminal Wigner's friend thought experiment, for which we explicitly show that decoherence is not capable of destroying macroscopic quantum effects. Instead, even under full decoherence of Wigner's friend---which from the point of view of the standard experiment destroys all quantum effects---a careful information-theoretic analysis reveals residual entanglement. This entanglement, as we show, depends crucially on the information held by Wigner's friend and can thus only be revealed when the observer is taken into account and giving a quantum description. This sheds new light on a classic experiment and demonstrates that decoherence while recovering a semi-classical description, does not destroy all quantum effects at the macroscopic scale. Interestingly, the observer's information, and thus the entanglement, can be transmitted via classical channels~\cite{Deutsch2000}, which has important consequences for the security analysis of quantum key distribution protocols. For experiments testing the limits of quantum theory, this means that quantum effects may sometimes be hiding in plain sight.

\section{Methods}
Here we provide additional details and proofs of the theory results in the main text. We first discuss the relevance of Alice's information for the observed entanglement. We then prove that Eq.~\eqref{witcond} represents a valid witness and discuss the equivalence of this witnessing to a semi-classical approach. Finally, we prove that entanglement is observed independent of the measure that is used. All data is available from the authors upon reasonable request.

\subsection*{Experimental implementation}
We implemented the modified Wigner's friend experiment shown in Fig.~\ref{fig:Scheme} using the setup of Fig.~\ref{fig:Setup}. Specifically, we use two Sagnac-based photon-pair sources, pumped by a \SI{1.6}{ps} pulsed laser at \SI{775}{nm} to produce single photons at $\SI{1550}{nm}$ via collinear type-II spontaneous parametric down-conversion in a \SI{22}{mm} long periodically-poled KTP (PPKTP) crystal. Quantum information is encoded in the polarization degree of freedom of these photons, with $\ket{0}$ and $\ket{1}$ represented by horizontal and vertical polarization, respectively. We only employ soft spectral filtering with a bandwidth of \SI{3}{nm} to reach a source brightness of about $\sim 3000$ pairs/mW/s at a heralding efficiency of $\sim 55\%$. 
One of these sources is pumped with diagonally polarized light to generate entangled states $\ket{\Psi_}$ with a typical fidelity of $\mathcal{F}=99.62^{+0.01}_{-0.04}\%$. The other source is pumped with linearly polarized light and additionally polarization filtered, to produce separable states $\ket{+}\ket{0}$. The latter of these photons is immediately detected to serve as a heralding signal for the experiment, using superconducting nano-wire single-photon detectors (SNSPDs) with a detection efficiency $\sim 80\%$ and a time-tagging module with a temporal resolution of \SI{156}{ps}.

One photon from the entangled pair then undergoes non-classical interference with the remaining single photon in state $\ket{+}$ at a polarizing beam splitter, with an interference visibility of $91.80^{+1.73}_{-1.73} \%$. This implements a so-called type-I fusion gate, and leaves the three photons in modes $\vs{atm}$ in the initial state of the experiment, Eq.~\eqref{initial}. A tunable amount of dephasing is then introduced to the state by means of random phase flips imparted by a HWP in mode $\vs{m}$ in a fraction of the experimental runs determined by the dephasing strength. In case of full dephasing, this prepares the state of Eq.~\eqref{atm}.

Alice then performs a projective measurement on mode ${\vs m}$, with the result recorded in the time-tagging module. Subsequently, Wigner performs a Bell-state measurement (BSM) the photons in modes $\vs a$ and $\vs t$. Using quantum tomography we calibrated the fidelity of this measurement to be $\mathcal{F}_{\textsc{bsm}}=96.84^{+0.05}_{-0.05}$. In order to evaluate the witnesses of Eq.~\eqref{witness}, Wigner then classically combines his BSM result with the result of Alice's measurement. Finally, we can test the importance of Alice's information by implementing her measurement in a sub-optimal manner, using non-orthogonal projectors, as described in the main text.

\subsection*{Relevance of Alice's information}
Here we consider a slightly more general scenario than in the main text, where instead of Eq.~\eqref{initial}, the initial state is given by
\begin{equation}
\label{initial_p}
\ket{\Phi}_{\vs{atm}} = \sqrt{p}\ket{\Psi_+}_{\vs{at}}\ket{0}_{\vs m} {+} \sqrt{1-p}\ket{\Psi_-}_{\vs{at}}\ket{1}_{\vs m}.
\end{equation}
After decoherence the state from Alice's point of view is thus given by
\begin{equation}
\hat{\varrho}_{\vs {atm}}^{(\vs A)} = p\ket{\Psi_+}_{\vs{at}}\bra{\Psi_+} {\otimes} \ket{0}_{\vs m}\bra{0}
+ (1-p) \ket{\Psi_-}_{\vs{at}}\bra{\Psi_-} {\otimes} \ket{1}_{\vs m}\bra{1}.
\end{equation}
Consequently, Wigner's state will take the form
\begin{equation}\label{atAWp}
\hat{\varrho}_{\vs{atA}}^{(\vs W)} = p\ket{\Psi_+}_{\vs{at}}\bra{\Psi_+}{\otimes} \hat{\tau}_{\vs A} + (1-p)\ket{\Psi_-}_{\vs{at}}\bra{\Psi_-}{\otimes} \hat{\upsilon}_{\vs A}.
\end{equation}
As we will show in the following, the parameter $p$ in some sense determines the importance of Alice's information for the final entanglement, as measured by Wigner. This is complementary to the argument in the main text, where we studied the effect of imperfect distinguishability of Alice's results. It is easy to see from Eq.~\eqref{initial_p} that the case $p=1/2$ recovers the scenario in the main text, whereas in the cases $p{=}0$ and $p{=}1$ the presence of Alice is irrelevant to the two-qubit entanglement emerging in the partition $(\vs a|\vs t)$. We now show that the parameter $p$ indeed measures the significance of Alice's information to Wigner. To see this, we trace out Alice from Wigner's state in Eq.~\eqref{atAWp} to obtain
\begin{equation}
\hat{\varrho}_{\vs{at}}^{(\vs W)} (p) = p |\Psi_+\rangle_{\vs{at}}\langle\Psi_+| + (1-p) |\Psi_-\rangle_{\vs{at}}\langle\Psi_-|.
\end{equation}  
It is now straightforward to calculate the negativity of Wigner's state without accessing Alice's information, $\hat{\varrho}_{\vs{at}}^{(\vs W)}$, as
\begin{equation}
\label{EqNoWitness}
E_N[\hat{\varrho}_{\vs {at}}^{(\vs W)}(p);({\vs a|\vs{t}})] = \left|\frac{1}{2}-p\right|.
\end{equation}
Thus, we see that disregarding Alice costs more to Wigner as $p$ tends to a half, in the sense that, while for $p=0,1$ he obtains the maximum distillable entanglement between the two qubits, for $p=1/2$ he obtains no distillable entanglement. Interestingly, just accessing Alice's information (assuming perfect distinguishability of $\hat{\tau}_{\vs A}$ and $\hat{\upsilon}_{\vs A}$) eliminates the $p$ dependence so that the distillable entanglement reaches its maximum as $E_N[\hat{\varrho}_{\vs {atA}}^{(\vs W)}(p);({\vs a|\vs{tA}})]=1/2$ for all values of $p$.

\subsection*{Proof of witnesses in Eq.~\eqref{witness}}
We aim to show that the two witnesses in Eq.~\eqref{witness} are able to detect entanglement of the state of Eq.~\eqref{atAW}. We first notice that both $\hat{W}_{1}$ and $\hat{W}_{2}$ have nonnegative expectation values over all separable states within $(\vs a|\vs{tA})$ and $(\vs {aA}|\vs t)$ partitions. Therefore, a negative expectation value signals the entanglement of the state under consideration in both partitions. 

We now show that any nonzero distinguishability of Alice's states in response to the question ``up or down?'' implies the negativity of the witness values.
Let us first calculate the traces ${\rm Tr}\hat{\varrho}_{\vs {{atA}}}^{(\vs W)}\hat{W}_{1}$ and ${\rm Tr}\hat{\varrho}_{\vs {atA}}^{(\vs W)}\hat{W}_{2}$ as below:
\begin{equation}
\begin{split}
{\rm Tr}\hat{\varrho}_{\vs {atA}}^{(\vs W)}\hat{W}_{1} = \frac{1}{4}(\tau_{\rm u} - \upsilon_{\rm u}),\\
{\rm Tr}\hat{\varrho}_{\vs {atA}}^{(\vs W)}\hat{W}_{2} = \frac{1}{4}(\tau_{\rm d} - \upsilon_{\rm d}),
\end{split}
\end{equation}
where we have used the facts that ${\rm Tr}\left[(\ket{\Phi_+}_{\rm at}\bra{\Phi_+}^{\mathsf{T}_{\vs t}})(\ket{\Psi_\pm}_{\vs {at}}\bra{\Psi_\pm})\right] = \pm \frac{1}{2}$ (alternatively ${\rm Tr}\left[(\ket{\Phi_-}_{\vs {at}}\bra{\Phi_-}^{\mathsf{T}_{\vs t}})(\ket{\Psi_\pm}_{\vs {at}}\bra{\Psi_\pm})\right] = \mp \frac{1}{2}$), and defined $\rm{Tr}\left[\hat{\tau}_{\vs A}\ket{\text{``up''}}_{\vs A}\bra{\text{``up''}}\right] = \tau_{\rm u}$, $\rm{Tr}\left[\hat{\upsilon}_{\vs A}\ket{\text{``up''}}_{\vs A}\bra{\text{``up''}}\right] = \upsilon_{\rm u}$, and similarly, ${\rm Tr}\left[\hat{\tau}_{\vs A}\ket{\text{``down''}}_{\vs A}\bra{\text{``down''}}\right] = \tau_{\rm d}$ and ${\rm Tr}\left[\hat{\upsilon}_{\vs A}\ket{\text{``down''}}_{\vs A}\bra{\text{``down''}}\right] = \upsilon_{\rm d}$.
We also notice that $\tau_{\rm d}+\tau_{\rm u}=1$ and $\upsilon_{\rm d}+\upsilon_{\rm u}=1$ and thus, ${\rm Tr}\hat{\varrho}_{\vs {atA}}^{(\vs W)}\hat{W}_{1} = -{\rm Tr}\hat{\varrho}_{\vs {atA}}^{(\vs W)}\hat{W}_{2}$.
Consequently, the only way for both witnesses to result in nonnegative values is that
\begin{equation}\label{witcond}
\begin{split}
\tau_{\rm u} = \upsilon_{\rm u},\\
\tau_{\rm d} = \upsilon_{\rm d},
\end{split}
\end{equation}
which means the two states $\hat{\tau}_{\vs A}$ and $\hat{\upsilon}_{\vs A}$ are indistinguishable in response to the question ``up or down?''.
Importantly, this can be true only if $\hat{\tau}_{\vs A}=\hat{\upsilon}_{\vs A}$, or $\hat{\tau}_{\vs A}$ and $\hat{\upsilon}_{\vs A}$ have unequal off-diagonal elements. The latter implies that Alice's state contains some coherence with respect to the basis $\{\ket{\text{``up''}}_{\vs A},\ket{\text{``down''}}_{\vs A}\}$, i.e., it is in a superposition of ``up'' and ``down''.

Finally, it is worth noticing that witness operators $\hat{W}_{1}$ and $\hat{W}_{2}$ together with the witnesses $\hat{W}'_{1}{=}|\Phi_-\rangle_{\rm at}\langle\Phi_-|^{\mathsf{T}_{\rm t}}{\otimes} |\text{``up"}\rangle_{\rm A}\langle \text{``up"}|$ and $\hat{W}'_{2}=|\Phi_-\rangle_{\rm at}\langle\Phi_-|^{\mathsf{T}_{\rm t}}{\otimes} |\text{``down"}\rangle_{\rm A}\langle \text{``down"}|$ are sufficient for detecting the entanglement of the more general form of state $\hat{\varrho}_{\rm atA}^{(\rm W)}(p)$ in Eq.~\eqref{atAWp} for any value of $p$ different from $1/2$.

\subsection*{Equivalence of witnessing and the semi-classical approach}
It is important to note that the witnessing procedure described in the main text can be thought of as the equivalent to the semi-classical approach via the assignments $\text{``up?''} \leftrightarrow \hat{\Pi}_{\rm u}=|\text{``up''}\rangle_{\vs A}\langle \text{``up''}| $ and $\text{``down?''} \leftrightarrow \hat{\Pi}_{\rm d}=|\text{``down''}\rangle_{\vs A}\langle \text{``down''}|$.
Consequently, the \emph{no witnessing} condition of Eq.~\eqref{EqNoWitness} simply reduces to the condition on the distance between classical probability distributions $p_{\rm u/d}={\rm Tr}\hat{\Pi}_{\rm u/d}\hat{\tau}_{\vs A}$ and $q_{\rm u/d}={\rm Tr}\hat{\Pi}_{\rm u/d}\hat{\upsilon}_{\vs A}$,
\begin{equation}
\frac{1}{2}\sum_{x={\rm u,d}}|p_x-q_x| = \|p-q\|_1 = 0.
\end{equation}
Now, to further make sense of the fact that the semi-classical description only gives a lower bound on the amount of distillable entanglement in our protocol, we recall the equivalent statement that \emph{the value by which a witnessing inequality is violated only puts a lower bound on the amount of entanglement within the quantum state}~\cite{Brandao2005,Shahandeh2017}.

\subsection*{Wigner's observed entanglement is measure-independent}
\paragraph*{Theorem.}
Suppose that $\hat{\varrho}_{\vs {XYZ}}{=}\sum_i p_i \hat{\sigma}_{{\vs {XY}};i}{\otimes}\hat{\sigma}_{{\vs Z};i}$ and $E$ is a convex measure of entanglement satisfying the strong monotonicity condition.
Then,
\begin{equation}
\label{EntEqv}
\begin{split}
\max p_iq_iE[\hat{\sigma}_{{\vs {XY}};i};({\vs X|\vs Y})] \leqslant &\\
E[\hat{\varrho}_{\vs {XYZ}};({\vs {XZ}|\vs Y})]=&E[\hat{\varrho}_{\vs {XYZ}};({\vs X|\vs{YZ}})]\\
\leqslant & \sum_i p_i E[\hat{\sigma}_{{\vs {XY}};i};({\vs X|\vs Y})],
\end{split}
\end{equation}
where $({\vs {XZ}|\vs Y})$, $({\vs X|\vs{YZ}})$, and $(\vs X|\vs Y)$ represent different bipartitions of the subsystems.
The maximization is over the set of states $\hat{\sigma}_{{\vs {XY}};i}$ with nonzero entanglement for which $q_i$ represents the probability of distinguishing the corresponding \emph{shield} state $\hat{\sigma}_{{\vs Z};i}$ within the set $\{\hat{\sigma}_{{\vs Z};i}\}$.
The upper bound is satisfied if $\{\hat{\sigma}_{{\vs Z};i}\}$ are unambiguously distinguishable.

\textit{Proof.} First, suppose that the shield state $\hat{\sigma}_{{\vs Z};J}$ can be distinguished from elements of the set $\{\hat{\sigma}_{{\vs Z};i}\}$ with probability $q_J$.
Then, assume that the corresponding bipartite state $\hat{\sigma}_{{\vs {XY}};J}$ is entangled.
By monotonicity of $E$, it is then true that $0 < p_Jq_JE[\hat{\sigma}_{{\vs {XY}};J};(\vs X|\vs Y)] \leqslant E[\hat{\varrho}_{\vs {XYZ}};({\vs {XZ}|\vs{Y}})]$; otherwise $\vs X$ and $\vs Y$ could increase their shared entanglement via local operations and classical communication (LOCC) by measuring system $\vs Z$ and postselecting on index $J$.
Thus, the first inequality follows.

Second, the convexity of $E$ implies that $E[\hat{\varrho}]\leqslant \sum_i p_i E[\hat{\sigma}_i]$ for every decomposition of the state $\hat{\varrho}=\sum_i p_i \hat{\sigma}_i$.
Using the fact that $E[\hat{\sigma}_{{\vs {XY}};i}\otimes\hat{\sigma}_{{\vs Z};i};({\vs {XZ}|\vs{Y}})]=E[\hat{\sigma}_{{\vs {XY}};i}\otimes\hat{\sigma}_{{\vs Z};i};({\vs X|{\vs{YZ}}})]=E[\hat{\sigma}_{{\vs {XY}};i};({\vs X|\vs{Y}})]$, the second inequality follows.

To show the saturation condition, again, we use the strong monotonicity, that is $E[\hat{\varrho}]\geqslant \sum_i q_i E[\hat{\sigma}_i]$ where each $\hat{\sigma}_i$ is obtained from $\hat{\varrho}$ via LOCC with probability $q_i$.
Combining this with the convexity property, we have $E[\hat{\varrho}]= \sum_i p_i E[\hat{\sigma}_i]$ if $\hat{\varrho}=\sum_i p_i \hat{\sigma}_i$ and there exists some LOCC such that $\hat{\varrho}\xrightarrow{\rm LOCC}\hat{\sigma}_i$ with probability $p_i$.
Now, it is evident that if $\hat{\varrho}_{\vs {XYZ}}=\sum_i p_i \hat{\sigma}_{{\vs {XY}};i}\otimes\hat{\sigma}_{{\vs Z};i}$ and the system $\vs Z$ is being held by $\vs X$ or $\vs Y$, that is within bipartitions $({\vs {XZ}|\vs{Y}})$ and $({\vs X|\vs{YZ}})$, then, $\{\hat{\sigma}_{{\vs Z};i}\}$ being unambiguously distinguishable (i.e., having disjoint supports) implies that $\hat{\varrho}_{\vs {XYZ}}\xrightarrow{\rm LOCC}\hat{\sigma}_{{\vs {XY}};i}\otimes\hat{\sigma}_{{\vs Z};i}$ with probability $p_i$; it is sufficient that $\vs X$ or $\vs Y$ make a measurement on $\vs Z$ and unambiguously determine the index $i$.
Hence, $E[\hat{\varrho}_{\vs {XYZ}};({\vs {XZ}|\vs{Y}})]=E[\hat{\varrho}_{\vs {XYZ}};({\vs X|\vs{YZ}})]=\sum_i p_i E[\hat{\sigma}_{{\vs {XY}};i}\otimes\hat{\sigma}_{{\vs Z};i};({\vs {XZ}|\vs{Y}})]$.
\qed

Considering now $\vs Z {\rightarrow} \text{Alice}$ in the theorem above we find that even if there is only one entangled state within the ensemble $\{\hat{\sigma}_{{\vs {XY}};i}\}$, say $\hat{\sigma}_{{\vs {XY}};J}$, that can be distilled with a nonzero probability $q_J$ by distinguishing the corresponding state of Alice $\hat{\sigma}_{{\vs A};J}$ among $\{\hat{\sigma}_{{\vs A};i}\}$, then the subsystem Alice-$\vs X$ ($\vs Y$) is entangled to the subsystem $\vs Y$ ($\vs X$).

\begin{acknowledgments}
This project was supported by the Australian Research Council Centre of Excellence for Quantum Computation and Communication Technology (CE170100012) and the UK Engineering and Physical Sciences Research Council (grant number EP/N002962/1). FS acknowledges support from the Royal Commission for the Exhibition of 1851. FC acknowledges support through an Australian Research Council Discovery Early Career Researcher Award (DE170100712). MR acknowledges funding from the European Union's Horizon 2020 research and innovation programme under the Marie Sk\l{}odowska-Curie grant agreement No 801110 and the Austrian Federal Ministry of Education, Science and Research (BMBWF). FG acknowledges studentship funding from EPSRC under grant no. EP/L015110/1. This publication was made possible through the support of a grant from the John Templeton Foundation. The opinions expressed in this publication are those of the authors and do not necessarily reflect the views of the John Templeton Foundation. We acknowledge the traditional owners of the land on which the University of Queensland is situated, the Turrbal and Jagera people.
FS conceived the project. FS, FC, and AL developed the theory. MR, FS, and AF conceived the experiment. MR, MP, FG, PB, AP and DK performed the experiment, and MR analyzed the data. TR and AF supervised the project. All authors contributed to writing the manuscript.
\end{acknowledgments}

\bibliography{MacroEntanglement.bib}
\bibliographystyle{apsrev4-1new}

\end{document}